# Tapered MMI splitters with unconstrained splitting ratio on a thick SOI platform

Matteo Cherchi*, Mikko Harjanne, Katherine Bryant, Fei Sun, Päivi Heimala, and Timo Aalto
VTT Technical Research Centre of Finland Ltd, Tietotie 3, 02150 Espoo, Finland

## ABSTRACT

We have systematically studied multimode interferometer (MMI) splitters made from multiple tapered sections. The goal is to create a library of robust and low-loss splitters covering all splitting ratios (SR) for our silicon photonics platform based on 3 µm thick waveguides. The starting point is always a non-tapered canonical MMI either with general symmetry (canonical SRs 50:50, 100:0, and reciprocal ratios), with mirror symmetric restricted symmetry (canonical SRs 85:15, 50:50, 100:0, and reciprocal ratios), and with point-symmetric restricted symmetry (canonical SRs 72:28 and 28:72). Splitters of these three types are then divided into one to four subsections of equal length, leading to 12 possible different configurations. In each of these subsections, the width is first linearly tapered either up or down and then tapered back to its starting value ensuring mirror symmetry. For all twelve configurations, we carried out an extensive campaign of numerical simulations. For each given width change, we scanned the splitter length and calculated the power in the fundamental mode at the output as well as its relative phase. We then selected the designs with sufficiently low loss and mapped their SR as a function of either the change in width change or length, therefore creating systematic maps for the design of MMI splitters with any SR. Eventually, we selected and fabricated a subset of designs with SRs ranging from 5:95 to 95:5 in steps of 5% and validated their operation through optical measurements.

## 1   INTRODUCTION

Silicon photonics platforms based on a micron-scale thick device layer[1,2] allow for a unique combination of tight bends[3–5] and relatively large modes, enabling dense integration with low propagation losses (down to 3 dB/m demonstrated[6]) and low coupling losses over a very broad wavelength range. One of the main challenges in these platforms is the realization of compact power splitters. The high mode confinement in the strip waveguide makes coupling to adjacent waveguides very difficult, meaning that in practice, only directional couplers made from rib-waveguides are possible due to fabrication constraints. On the other hand, rib directional couplers are typically several millimetres long, due to the low lateral index contrast and to the large bending radii in the input and output sections. Even more important, etch-depth variations of around ±5% at wafer scale result in major variability of the SRs around the wafer.

For these reasons, strip MMIs are the most popular choice as power splitters. Thanks to the strong light confinement, they operate very close to the ideal analytical model[7], also resulting in low losses. Furthermore, strip MMIs have a high tolerance to fabrication variations and enable a broadband operation  which can be outperformed only by the significantly larger-footprint adiabatic couplers[8]. Unfortunately, canonical MMIs come with a limited set of SRs, which is a strong limitation, e.g., for the design of lattice filters. In the past, we have overcome this limitation by using double MMIs, either with tapered waveguide connections[9] or with bent waveguide connections[10]. The main limitation of this approach is that, even though MMIs are very robust to fabrication variations, in general the phase shifters are not as robust. Furthermore, by construction, the double MMIs are more than twice the length of single MMIs and, if the phase shifters are implemented through bent sections (which proved to work better than the tapered waveguides), their layout on mask is complicated by the fact that the output waveguides are in general not collinear with the input waveguides.

## 2   TAPERED MMIS

A more robust approach is using tapered MMIs, as proposed by Besse et al.[11] back in 1996, and later systematically studied by Doménech et al.[12], based on submicron silicon waveguides. By suitably tapering up or down any of the canon-ical 50:50, 85:15 or 72:28 MMIs, it is possible to achieve any desired SR with low loss, broadband operation, and high tolerance to fabrication variations. Given that we already have successfully demonstrated the use of tapered MMIs on a thick SOI platform[13], we decided to study them systematically to identify the best configurations and possibly add them to the process design kit of the platform. With this goal in mind, we launched a semi-automated simulation campaign based on the eigenmode expansion method (with the commercial software FimmProp by Photon Design) exploring a very broad range of possible configurations, as summarised in Figure 1.

*matteo.cherchi@vtt.fi; phone +358 40 684 9040, ORCID 0000-0002-6233-4466, www.vtt.fi

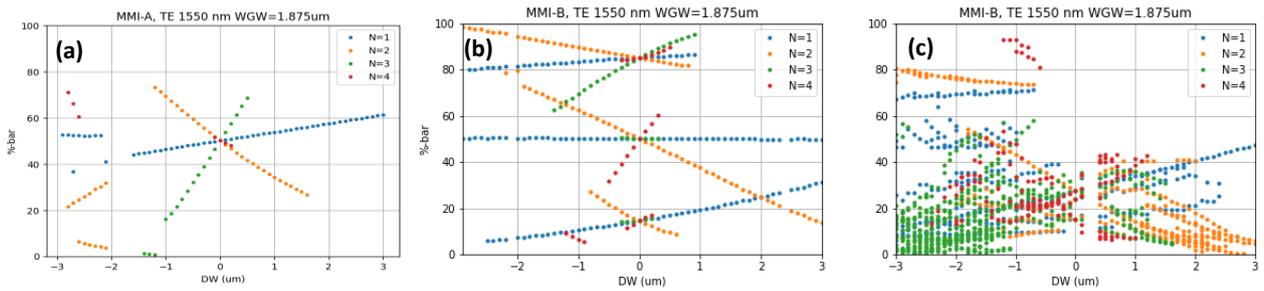

Figure 1. Summary of the space of configurations we have explored by systematic simulations.

There we have labelled MMIs: with general symmetry as Type A, with restricted mirror symmetry as Type B, and with restricted point symmetry as Type C. The SR of each of them can be changed from the canonical working points (the first number in the SR represents the power fraction in the cross port) by linear tapers with $N = 1…4$ periods, either tapering up (width change $\Delta W > 0$) or down ($\Delta W < 0$). We have retained only the configurations with loss below 5% (i.e., less than 0.225 dB loss) and discarded the other ones. A summary of the found solutions is presented in Figure 2 for all three types of MMIs, highlighting that all possible SRs can be reached one way or the other.

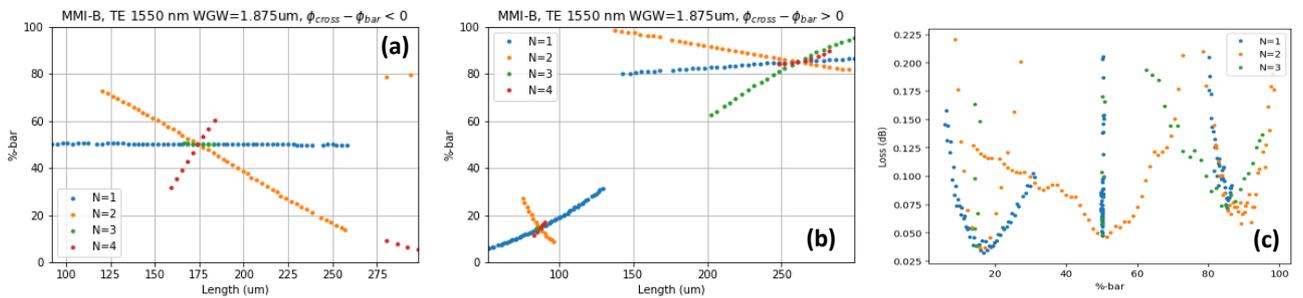

Figure 2. Bar port power fraction versus $\Delta W$ for the found low loss solutions of Type A, B, and C with $N = 1…4$.

In this work, we focus only on Type B MMIs for the experimental verification of the simulated results, as they alone can cover almost the full range of SRs. Remarkably, Figure 2.b shows that for $N = 1$, and $N = 3$, tapering does not affect the SR around the 50:50 working point. In Figure 3, we show the lengths associated with Type B solutions, where Figure 3.a corresponds to negative phase difference between the cross and the bar outputs ports (i.e., around -90°) and Figure 3.b to positive values (around +90°).

Figure 3. Simulation maps showing a) the length of the found solutions with negative relative output phase; b) same corresponding to positive relative output phase; and c) associated distribution of losses ($N = 4$ is missing).

In Figure 3.c we also show the loss associated with different solutions. For the sake of clarity, in this last plot we have excluded the least interesting case $N = 4$, corresponding to steeper slopes, meaning lower tolerance to fabrication variations.

## 3  EXPERIMENTAL RESULTS

We have devised a simple algorithm that picks up the shortest and lowest loss Type B solution (with $N = 1…3$) for any given SR and used it to draw nineteen tapered MMIs covering SRs from 95:05 to 05:95 in steps of 5%. The devices were fabricated in a multi-process run using our standard passive module, which includes etching of the facets and anti-reflection coating deposition all at the wafer scale. We have measured the transmission of all devices from a single chip and characterized the spectral response of both output ports (top port labelled 1 and bottom port labelled 2) corresponding to both input ports (labelled following the same convention). Power fractions $T_{io}$ (where $i$ and $o$ represent the input and output port numbers respectively) of eight of the nineteen measured splitters are shown in Figure 4 as a function of wavelength and are normalized to the overall transmitted power. Minimum losses (normalized to a straight waveguide transmission spectrum, not plotted) vary from less than 0.1 dB to up to 0.5 dB, depending on the device. Furthermore, in some of the devices, we noticed a significant variation of the minimum loss depending on the chosen input, despite the fact that all designs should be perfectly symmetric. We ascribe such variations to possible defects on the input and output facets. Apart from the 60:40 and the 50:50 splitters, all the other splitters have been found on target within ±1% accuracy. Possible reasons for the inaccuracy of those two cases are still under investigation. Both designs have $N = 2$ and $\Delta W > 0$ (like the 55:45 splitter in Figure 4.e).

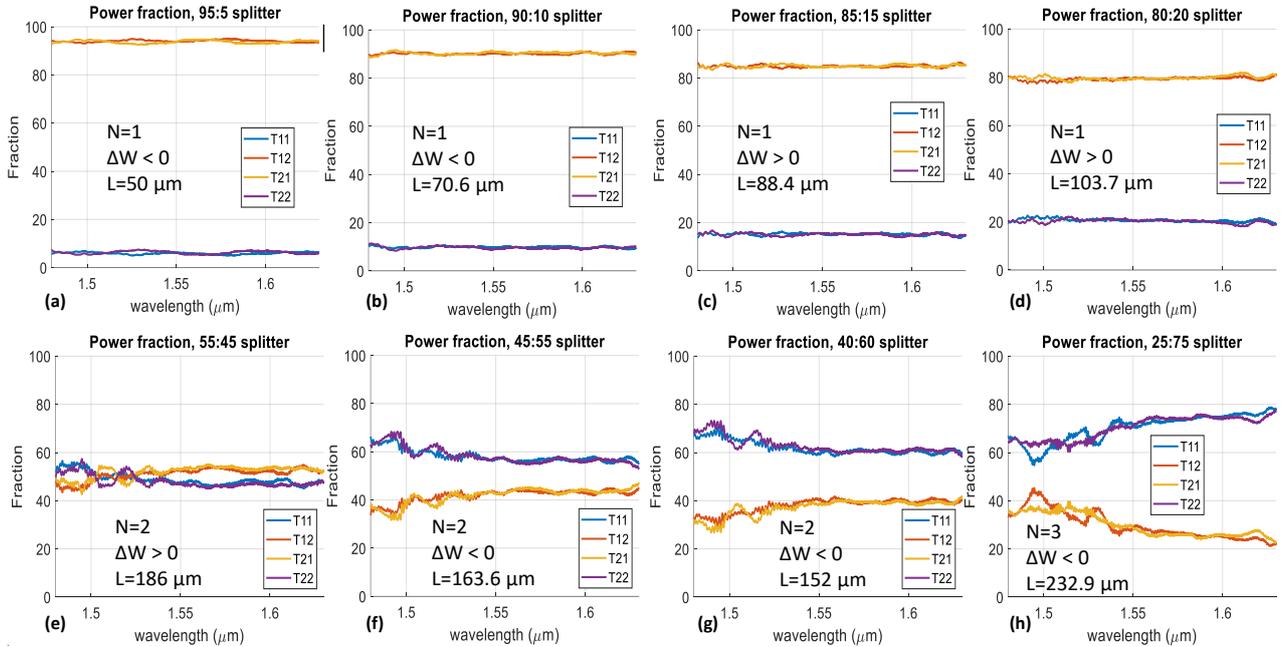

Figure 4. Power fraction vs wavelength for eight of the designed splitters. For each design we show the number of periods, the sign of the tapering and the length of the device.

## 4  CONCLUSIONS

Starting from the main three known layouts and their canonical working points, we have systematically simulated tapered MMIs with number of periods $N = 1…4$ and selected those instances showing losses smaller than 0.225 dB. The results clearly show that tapering allows us to achieve any desired SR. In particular, tapered MMI splitters with restricted mirror symmetry can alone cover almost the whole range. For experimental verification, we have chosen nineteen Type B tapered MMIs covering the range from 95:05 to 05:95 and fabricated them on our thick SOI platform. All the fabricated devices show low loss and relatively broadband operation, and seventeen of them successfully reached the targeted splitting ratio within ±1% accuracy. In future studies, we will investigate the non-ideal response of the two remaining designs as well as the asymmetric operation of some the splitters.

## 5 ACKNOWLEDGEMENTS

We acknowledge support by the Academy of Finland Flagship Programme, Photonics Research and Innovation (PREIN), decision number 320168, and by the Business Finland project PICAP, decision number 44065/31/2020.